\documentclass[prd, aps,]{revtex4}

\def\be{\begin{equation}}
\def\te{\end{equation}}
\def\ee{\end{equation}}
\def\ba{\begin{eqnarray}}
\def\bea{\begin{eqnarray}}
\def\nn{\nonumber\\}
\def\tea{\end{eqnarray}}
\def\ea{\end{eqnarray}}
\def\eea{\end{eqnarray}}

\newskip\humongous \humongous=0pt plus 1000pt minus 1000pt

\newif\ifdtup

\begin{document}


%
%

\title{Stochastic Gross-Pitaevsky Equation for BEC
via Coarse-Grained Effective Action  }

\author{Esteban Calzetta\footnote{calzetta@df.uba.ar }}

\address{{\it {\small Departamento de Fisica, Facultad de Ciencias
Exactas y Naturales,}}\\
{\it {\small Universidad de Buenos Aires- Ciudad Universitaria, 1428
Buenos Aires, Argentina}}}

\author{B. L. Hu\footnote{blhu@umd.edu}}

\address{{\it {\small Department of Physics, University of Maryland,
College Park, MD 20742, U.S.A.}}}

\author{Enric Verdaguer\footnote{enric.verdaguer@ub.edu }}

\address{{\it {\small Departament de Fisica Fonamental and Institut
de Ciencies del Cosmos,}}\\
{\it {\small Universitat de Barcelona, Av.~Diagonal 647, 08028
Barcelona, Spain}}}

\date{February 2, 2007}


\centerline{\it Dedicated to Professor Bambi Hu in honor of his 60th
birthday, with appreciation and love -- Bei-Lok}
\bigskip
\begin{abstract}
We sketch the major steps in a functional integral derivation of a
new set of Stochastic Gross-Pitaevsky equations (GPE) for a
Bose-Einstein condensate (BEC) confined to a trap at zero
temperature with the averaged effects of non-condensate modes
incorporated as stochastic sources. The closed-time-path (CTP)
coarse-grained effective action (CGEA) or the equivalent influence
functional method is particularly suitable because it can account
for the full back-reaction of the noncondensate modes on the
condensate dynamics self-consistently. The Langevin equations
derived here containing nonlocal dissipation together with colored
and multiplicative noises are useful for a stochastic (as
distinguished from say, a kinetic) description of the nonequilibrium
dynamics of a BEC.  This short paper contains original research
results not yet published anywhere.
\end{abstract}
\maketitle


\section{Introduction}

Two major paradigms are commonly adopted in describing the
nonequilibrium dynamics of a many-body system. One is by way of
kinetic theory \cite{GarZolPRA,KinThyBEC} based on the Boltzmann
equation and its generalizations, which evolve the lowest order
correlation functions (in the BBGKY hierarchy). The other captures
the stochastic and dissipative dynamics of an open system, with the
environment functioning as sources of noise with specific features
determined from first principles, not put in by hand; and their
overall effect engendering dissipation in the open system dynamics.
Usually for a system well discernible from and whose effects
dominate over its environment (the condensate is an open system with
the noncondensate as its environment), the stochastic dynamics is
often an easier and perhaps cleaner approach. Thus explains the
great interest in deriving reliable stochastic equations for BEC
dynamics.

For BEC stochastic dynamics, historically Gardiner and his
co-workers \cite{GardinerSGPE} were the first to derive a set of
stochastic Gross-Pitaevsky equations. Their method combines the
kinetic theory and the open systems approaches. We present here an
alternative derivation via the closed-time-path (CTP) \cite{ctp}
coarse-grained effective action (CGEA) \cite{cgea} or the closely
related influence functional \cite{if} method.  This method has the
merit that it can account for the full back-reaction of the
noncondensate modes on the condensate dynamics self-consistently.
Our results are structurally similar to theirs but the content is
more complex: Our Langevin equations contains nonlocal dissipation
and colored and multiplicative noises. These two kernels
provide all the necessary information for a stochastic description
of the nonequilibrium dynamics of a BEC. \footnote{Similar efforts
were made along this line of thought for semiclassical stochastic
gravity, known as the Einstein-Langevin equation, and in quark gluon
plasma physics, known as the Bodeker equation.}

\section{Quantum Field Theory of BEC Dynamics}

For a field-theoretic description of BEC we begin with a second -
quantized field operator $\Psi \left( \vec x,t\right) $ which removes an
atom at the location $\vec x$ at times $t$. It obeys canonical
commutation relations
\begin{equation}
\left[ \Psi \left( \vec x,t\right) ,\Psi \left( \vec y,t\right) \right] =0,
\end{equation}
\begin{equation}
\left[ \Psi \left( \vec x,t\right) ,\Psi ^{\dagger }\left( \vec y,t\right)
\right] =\delta \left( \vec x-\vec y\right) . \label{etcr}
\end{equation}
The dynamics of this field is given by the Heisenberg equations of
motion (we choose units such that $\hbar =1$)
\begin{equation}
-i \frac{\partial }{\partial t}\Psi =\left[ \mathbf{H},\Psi
\right],
\end{equation}
where $\mathbf{H}$ is the Hamiltonian. The theory is invariant under
a global phase change of the field operator
\begin{equation}
\Psi \rightarrow e^{i\theta }\Psi ,\qquad \Psi ^{\dagger
}\rightarrow e^{-i\theta }\Psi ^{\dagger }.  \label{global}
\end{equation}
The constant of motion associated with this invariance through
Noether's theorem is the total particle number.

We shall consider only the simplest Hamiltonian
\begin{equation}
\mathbf{H}=\int d^d\vec x\;\left\{ \Psi ^{\dagger }H\Psi +\frac U2\Psi
^{\dagger 2}\Psi ^2\right\},  \label{nbodyh}
\end{equation}
where we have reduced the atom - atom interactions to a single
contact potential with a delta-like repulsion term. The coupling
constant $U$ is often parameterized in terms of the scattering
length $a$ and the mass $M$ of the atom as $U=4\pi a/M.$ The
single - particle Hamiltonian $H$ is given by
\begin{equation}
H\Psi =-\frac{1}{2M}\nabla ^{2}\Psi +V_{trap}\left(
\vec x\right) \Psi, \label{sparth}
\end{equation}
$V_{trap}\left( \vec x\right) $ denotes a confining trap potential. Then
the Heisenberg equation of motion
\begin{equation}
i \frac{\partial }{\partial t}\Psi =H\Psi +U\Psi ^{\dagger
}\Psi ^{2}, \label{Heisenberg}
\end{equation}
is also the classical equation of motion derived from the action
\begin{equation}
S=\int dt\:d^{d}x\;i \Psi ^{*}\frac{\partial }{\partial t}\Psi -\int dt\;%
\mathbf{H}.  \label{action}
\end{equation}

\section{Condensate as open system}

We adopt the quantum open system framework and define our system of
condensate as comprising a superposition of only a few low lying
modes of the one-particle Hamiltonian. The condensate while evolving
can excite the quantum fluctuations in the higher modes, which in
turn back-react on the condensate, thus modifying its test-field
dynamics (i.e, that which is obtained without backreaction). We
regard the condensate as an open system living in the environment
provided by the higher modes and try to incorporate the effects of
the coarse-grained environment in some way. The inclusion of
backreaction effects in our experience is best performed via the
coarse grained effective action or the  influence functional as they
ensure full self-consistency.

We first partition the total wave function into the system sector
and an environment sector. We do this by way of projection
operators.  Assume a complete set of one-particle states
\begin{equation}
H\psi_{\alpha}\left(\vec
x\right)=\omega_{\alpha}\psi_{\alpha}\left(\vec x\right).
\end{equation}
Choose an appropriate partition frequency  $\omega_{C}$ and define
two projectors
\begin{equation}
{\cal P}=\sum_{\omega_{\alpha}\leq\omega_C}\psi_{\alpha}\left(\vec
x\right)\psi_{\alpha}\left(\vec x'\right),
\end{equation}
which projects onto the low lying modes (band) comprising the
condensate, our system,  and
\begin{equation}
{\cal Q}=\delta\left(\vec x-\vec x'\right)-{\cal P}.
\end{equation}
which projects onto the sector of higher modes, the noncondensate,
constituting an environment for the system.  Define the condensate
band field operator
\begin{equation}
\phi \left( t,\vec x \right)={\cal P}\Psi \left( t,\vec x \right),
\end{equation}
and split \be \Psi \left( t,\vec x \right) = \phi \left( t,\vec x
\right) + \chi \left( t,\vec x \right). \ee Here, $\phi$ denotes the
{\it quantum} wave function of the condensate, not a c-number wave
function usually assumed.

We now split the action into condensate ($C$), non-condensate ($NC$)
and interaction ($int$) parts \be
S\left[\phi+\chi\right]=S_{C}\left[\phi\right]+
S_{NC}\left[\chi\right]+S_{int}\left[\phi,\chi\right].
\ee The first two are just the full action evaluated at the
corresponding field. $S_{int}$ may be written as the sum of three
terms in successive higher powers of the noncondensate operator: \be
S_{int}=S_{1}+S_{2}+S_{3}, \ee with \be S_1=-U\int dt\:d^{d}\vec
x\;\left\{\phi^{\dagger 2}\phi\chi +\chi^{\dagger
}\phi^{\dagger}\phi^{2}\right\}, \ee \be S_2=-\frac{U}{2}\int
dt\:d^{d}\vec x\;\left\{\phi^{\dagger 2}\chi^{2}+\chi^{\dagger
2}\phi^{2} +4\phi\phi\chi^{\dagger }\chi\right\}, \ee \be S_3=-U\int
dt\:d^{d}\vec x\;\left\{\phi^{\dagger}\chi^{\dagger }\chi^{2}
+\chi^{\dagger 2}\chi\phi\right\}, \ee where $\phi^\dagger$ is the
hermitian conjugated field of $\phi$. To simplify the appearances we
represent the pair $\left(\phi ,\phi^{\dagger}\right)$ as the up and
down components (respectively) of a spinor $\phi^{a}$, $a=1,2$, and
adopt the DeWitt convention where the space-time arguments are
assumed to be included in the spinor indices.

\section{CGEA and Stochastic Gross-Pitaevsky (Langevin) Equation}

We now introduce the coarse-grained effective action.  Assume for
simplicity that at $t=0$ the quantum state of the gas is described
by a factorizable density matrix \be
\rho\left[\phi^{1}+\chi^{1},\phi^{2}+\chi^{2},0\right]=
\rho_{C}\left[\phi^{1},\phi^{2},0\right]\rho_{NC}\left[\chi^{1},\chi^{2},0\right].
\ee The expectation values of field operators may be derived from
the usual closed time-path (CTP) generating functional \be
exp\left\{iW\left[J^{1},J^{2}\right]\right\}=\int
D\Psi^{1}\:D\Psi^{2}\;e^{i\left(S\left[\Psi^{1}\right]-S\left[\Psi^{2}\right]+\int
dt\:d^{d}x\;\left[J^{1}\Psi^{1}-J^{2}\Psi^{2}\right]\right)}\rho\left(0\right).
\ee However, if we are interested in the expectation values of the
condensate band operators, then we only need to couple sources to
these fields, namely \be {\cal Q}J^{1,2}=0. \ee The integral over
non-condensate fields may be performed, and we obtain the condensate
generating functional \be
exp\left\{iW_{C}\left[J^{1},J^{2}\right]\right\}=\int
D\phi^{1}\:D\phi^{2}\;e^{i\left(S_{CGEA}\left[\phi^{1},\phi^{2}\right]+\int
dt\:d^{d}x\;\left[J^{1}\phi^{1}-J^{2}\phi^{2}\right]\right)}\rho_{C}\left(0\right),
\label{cgf} \ee where the coarse-grained effective action is  \be
S_{CGEA}\left[\phi^{1},\phi^{2}\right]=
S_{C}\left[\phi^{1}\right]-S_{C}\left[\phi^{2}\right]
+ S_{IF} \left[\phi^{1},\phi^{2}\right], \ee where $S_{IF}$ is
the influence action and \be exp\left\{i
S_{IF}\left[\phi^{1},\phi^{2}\right]\right\}=\int
D\chi^{1}\:D\chi^{2}\;e^{i\left(S_{NC}\left[\chi^{1}\right]+
S_{int}\left[\phi^{1},\chi^{1}\right]-S_{NC}\left[\chi^{2}\right]-
S_{int}\left[\phi^{2},\chi^{2}\right]\right)}\rho_{NC}\left(0\right).
\ee is the influence functional (IF).

To make the physical content of the CGEA more explicit, it is
convenient to introduce new field variables \be
\phi_+=\frac{1}{2}\left(\phi^{1}+\phi^{2}\right),
\;\;\phi_-=\phi^{1}-\phi^{2}. \ee If we Taylor expand
$S_{CGEA}$ in powers of $\phi_-$ we find that there is no
independent term, and that even terms are imaginary and odd terms
are real. The path integral (\ref{cgf}) admits a saddle point with
respect to $\phi_-$ at $\phi_-=0$. We concentrate on the
contribution of field configurations near this saddle point by
keeping only quadratic terms in $\phi_-$ in the CGEA \bea
S_{IF}\left[\phi_{+},\phi_{-}\right]&=&\int dt\:d^{d}\vec x\;{\cal
D}_{a}\left[\phi_+\right]\phi_-^{a}\nn &+&\frac{i}{2}\int
dt\:d^{d}\vec x\int dt'\:d^{d}\vec x'\;{\cal
N}_{ab}\left[\phi_+\right]\phi_-^{a}\phi_-^{b}+\ldots \tea To get an
explicit representation, we introduce the functional expectation
value \be \left\langle A\right\rangle\left[\phi_{+}\right]=\int
D\chi^{1}\:D\chi^{2}\;e^{i\left(S_{NC}\left[\chi^{1}\right]+
S_{int}\left[\phi_{+},\chi^{1}\right]-
S_{NC}\left[\chi^{2}\right]-
S_{int}\left[\phi_{+},\chi^{2}\right]\right)}A\rho_{NC}\left(0\right).
\ee Then \be {\cal D}_{a}\left[\phi_+\right]=\frac{\delta
S_{C}}{\delta\phi_+^{a}}+\frac{1}{2}\left\langle \frac{\delta
S_{int}}{\delta\phi_+^{a}}\left[\phi_+,\chi^{1}\right]+\frac{\delta
S_{int}}{\delta\phi_+^{a}}\left[\phi_+,\chi^{2}\right]\right\rangle\left[\phi_{+}\right],
\ee and \bea {\cal
N}_{ab}\left[\phi_+\right]&=&\frac{-i}{4}\left\langle
\frac{\delta^{2}
S_{int}}{\delta\phi_+^{a}\delta\phi_+^{b}}\left[\phi_+,\chi^{1}\right]-\frac{\delta^{2}
S_{int}}{\delta\phi_+^{a}
\delta\phi_+^{b}}\left[\phi_+,\chi^{2}\right]\right\rangle\left[\phi_{+}\right]\nn
&+&\frac{1}{4}\left\langle \left\{\frac{\delta
S_{int}}{\delta\phi_+^{a}}\left[\phi_+,\chi^{1}\right]+\frac{\delta
S_{int}}{\delta\phi_+^{a}}\left[\phi_+,\chi^{2}\right]\right\}\left\{\frac{\delta
S_{int}}{\delta\phi_+^{b}}\left[\phi_+,\chi^{1}\right]+\frac{\delta
S_{int}}{\delta\phi_+^{b}}
\left[\phi_+,\chi^{2}\right]\right\}\right\rangle\left[\phi_{+}\right]\nn
&-&\frac{1}{4}\left\langle \frac{\delta
S_{int}}{\delta\phi_+^{a}}\left[\phi_+,\chi^{1}\right]+\frac{\delta
S_{int}}{\delta\phi_+^{a}}
\left[\phi_+,\chi^{2}\right]\right\rangle\left[\phi_{+}\right]\left\langle
\frac{\delta
S_{int}}{\delta\phi_+^{b}}\left[\phi_+,\chi^{1}\right]+\frac{\delta
S_{int}}{\delta\phi_+^{b}}\left[\phi_+,\chi^{2}\right]\right\rangle\left[\phi_{+}\right].
\tea

{}From here one can deduce that generally a representation \be {\cal
N}_{ab}\left[\phi_+\right]=\sum_{ij}\nu_{ac,bd}^{ij}f^{c}_{i}
\left[\phi_+\right]f^{d}_{j}\left[\phi_+\right], \ee can be obtained
in terms of a set of local functionals $f^{a}_{i}$. This allows us
to rewrite the condensate generating functional as \bea
exp\left\{iW_{C}\left[J_+,J_-\right]\right\}&=&\int
D\xi^{i}_{ac}\:\Upsilon \left[\xi^{i}_{ac}\right]\;
D\phi_{+}\:D\phi_{-}\;e^{i\int dt\:d^{d}x\;J_-\phi_+}\nn && e^{i\int
dt\:d^{d}x\;\left\{{\cal D}_{a}\left[\phi_+\right]+
\sum_{i}\xi^{i}_{ac}f^{c}_{i}\left[\phi_+\right]+
J_+\right\}\phi_-^{a}}\rho_{C}\left(0\right), \label{cgf2} \tea
where the Feynman-Vernon trick of using a Gaussian functional
identity \cite{if} involving the stochastic distribution $\Upsilon$
is invoked. Here \be J_+=\frac{1}{2}\left(J^{1}+J^{2}\right),
\;\;J_-=J^{1}-J^{2}, \ee and the $\xi^{i}_{ac}$ are the Gaussian
stochastic sources with zero mean and cross correlation \be
\left\langle \xi^{i}_{ac}\xi^{j}_{bd}\right\rangle
=\nu_{ac,bd}^{ij}. \ee

It should be noted that the stochastic terms do not
appear explicitly, but only implicitly in the form of random contribution to the action, which then is averaged over.
However, performing the final integration over
$\phi_-^{a}$ we see that the only contributions to the path
integral come from condensate field configurations which
obey the equation \be {\cal P}\left\{{\cal
D}_{a}\left[\phi_+\right]+\sum_{i}\xi^{i}_{ac}f^{c}_{i}\left[\phi_+
\right]+J_+\right\}=0,\label{lantype}
\ee with random initial conditions weighted
by the Wigner function constructed from the initial density matrix
$\rho_{C}\left(0\right)$. The free evolution of the condensate is
obtained by setting $J_+=0$. In this sense, the condensate
field evolves as an ensemble of trajectories,
each obeying the stochastic Langevin-type equation (\ref{lantype}).
This is a consequence of the truncation of $S_{CGEA}$ to just quadratic terms in  $\phi_-^{a}$. Further terms in $S_{CGEA}$ would introduce effects
associated to the quantum nature of the condensate field.
The presence of the projection operator ${\cal P}$ in (\ref{lantype}) enforces the consistency of the system - environment split under time evolution.

We observe that because of the nonlocal
terms in ${\cal D}_{a}$ this equation is generally dissipative and
contain noises which are generally colored because of the
nonlocality of the cross correlations $\nu_{ac,bd}^{ij}$,  and
multiplicative because of nonlinearity in the $f^{c}_{i}$. There
ought to be a fluctuation-dissipation relation for each types of the
noise. A similar description for a theory with polynomial type of
system-environment coupling can be found in Hu, Paz and Zhang (1993)
of \cite{if}; see also \cite{refsugg}.

\section{Dissipation and Noise Kernels}

We can glean more physical meanings about dissipation and
fluctuations by trying to obtain more explicit expressions for these
kernels. We go back to our model of a gas confined in a
time-independent trap, and compute the CGEA to order $U^{2}$ at zero
temperature ($T=0$) for the vacuum. We obtain (from here on, to
simplify appearances, we drop the $+$ subindex on $\phi_+$) \bea
{\cal D}_{a}\left[\phi\right]&=&\frac{\delta
S_{C}}{\delta\phi^{a}}+\frac{iU^{2}}{2}\int^{t}dt'd\vec x'\nn
&&\left\{\frac{\partial\phi^{\dagger
2}\phi}{\partial\phi^{a}}\left(t,\vec x\right)\;F^{+}\left(t-t',\vec
x,\vec x'\right)\phi^{\dagger}\phi^{2}\left(t',\vec
x'\right)\right.\nn &-&
\frac{\partial\phi^{\dagger}\phi^{2}}{\partial\phi^{a}}\left(t,\vec
x\right)F^{-}\left(t-t',\vec x,\vec x'\right)\phi^{\dagger
2}\phi\left(t',\vec x'\right)\nn
&+&\frac{1}{2}\frac{\partial\phi^{\dagger
2}}{\partial\phi^{a}}\left(t,\vec x\right)F^{+2}\left(t-t',\vec
x,\vec x'\right)\phi^{2}\left(t',\vec x'\right)\nn
&-&\left.\frac{1}{2}\frac{\partial\phi^{2}}{\partial\phi^{a}}\left(t,\vec
x\right)F^{-2}\left(t-t',\vec x,\vec x'\right)\phi^{\dagger
2}\left(t',\vec x'\right)\right\}, \tea where \bea
F^{+}\left(t-t',\vec x,\vec x'\right)&=&\left\langle
\chi\left(t,\vec x\right)\chi^{\dagger }\left(t',\vec
x'\right)\right\rangle\nn &=&\sum_{\omega_{\alpha}>\omega_{C}}
e^{-i\omega_{\alpha}\left(t-t'\right)}\psi_{\alpha}\left(\vec
x\right)\psi_{\alpha}\left(\vec x'\right), \tea \bea
F^{-}\left(t-t',\vec x,\vec x'\right)&=&\left\langle
\chi\left(t',\vec x'\right)\chi^{\dagger }\left(t,\vec
x\right)\right\rangle\nn
&=&\sum_{\omega_{\alpha}>\omega_{C}}e^{i\omega_{\alpha}
\left(t-t'\right)}\psi_{\alpha}\left(\vec
x\right)\psi_{\alpha}\left(\vec x'\right). \tea With respect to the
noise terms, we find that they may be represented in terms of four
stochastic terms $\zeta^1$ to $\zeta^4$ \be
\xi^{i}_{ac}=\delta_{ac}\zeta^{i},\ i=1,\dots,4, \ee coupled to four
functions $f_1$ to $f_4$  \be
f_{1a}=\frac{1}{2}\frac{\partial}{\partial\phi^{a}}\left[\phi^{\dagger
2}\phi +\phi^{\dagger}\phi^{2}\right], \ee \be
f_{2a}=\frac{-i}{2}\frac{\partial}{\partial\phi^{a}}\left[\phi^{\dagger
2}\phi -\phi^{\dagger}\phi^{2}\right], \ee \be
f_{3a}=\frac{1}{2}\frac{\partial}{\partial\phi^{a}}\left[\phi^{\dagger
2} +\phi^{2}\right], \ee \be
f_{4a}=\frac{-i}{2}\frac{\partial}{\partial\phi^{a}}\left[\phi^{\dagger
2} -\phi^{2}\right]. \ee The nontrivial cross correlations are \be
\left\langle \zeta^{1}\left(t,\vec x\right)\zeta^{1}\left(t',\vec
x'\right)\right\rangle =\left\langle \zeta^{2}\left(t,\vec
x\right)\zeta^{2}\left(t',\vec x'\right)\right\rangle
=\frac{1}{2}F_{p}\left(t-t',\vec x,\vec x'\right), \ee \be
\left\langle \zeta^{1}\left(t,\vec x\right)\zeta^{2}\left(t',\vec
x'\right)\right\rangle =\frac{i}{2}F_m \left(t-t',\vec x,\vec
x'\right), \ee \be \left\langle \zeta^{3}\left(t,\vec
x\right)\zeta^{3}\left(t',\vec x'\right)\right\rangle =\left\langle
\zeta^{4}\left(t,\vec x\right)\zeta^{4}\left(t',\vec
x'\right)\right\rangle =\mathbf{F}_{p}\left(t-t',\vec x,\vec
x'\right), \ee \be \left\langle \zeta^{3}\left(t,\vec
x\right)\zeta^{4}\left(t',\vec x'\right)\right\rangle
=i\mathbf{F}_m\left(t-t',\vec x,\vec x'\right), \ee where \be
F_p=F^{+}+F^{-},\ \ F_m=F^{+}-F^{-},\ \ \mathbf{F}_p=F^{+2}+F^{-2},\
\ \mathbf{F}_m=F^{+2}-F^{-2}. \ee

The detailed forms of the noise and dissipation kernels in the
stochastic Gross-Pitaevsky (Langevin) equation are useful for a
comprehensive and in-depth study of the  stochastic  dynamics of the
condensate incorporating its interaction with the noncondensates.
Various aspects of this problem are understudy and results will be
reported elsewhere. \\

\noindent{\bf Acknowledgement} BLH thanks the organizers of this
symposium, Dr. Baowen Li and Dr. Leihan Tang, for their invitation
and hospitality.  This work is supported by the University of Buenos
Aires, CONICET and ANPCYT (Argentina);, the NSF-ITR (PHY-0426696),
NIST and NSA-LPS to the University of Maryland, and the Spanish MEC
Research Project No. FPA2004-04582.

\end{document}